\documentclass{amsart}
\usepackage{natbib}
\usepackage{graphicx}
\usepackage{amssymb}
\begin{document}
\begin{abstract}
The structure of the interaction Hamiltonian in the first order
$S-$matrix element of a Dirac particle in an Aharonov-Bohm (AB)
field is analyzed and  shown to have interesting  algebraic
properties. It is demonstrated that as a consequence of these
properties, this interaction Hamiltonian
 splits both the incident and outgoing waves in the the first order $S-$matrix into their
$\frac{\Sigma_3}{2}-$components ( eigenstates of the third
component of the spin). The matrix element can then be viewed as
the sum of two transitions taking place in these two channels of
the spin. At the level of partial waves, each partial wave of the
conserved total angular momentum is split into two partial waves
of the orbital angular momentum in a manner consistent with the
conservation of the total angular momentum quantum number.\\

\end{abstract}

\title[ Dirac Particle in an Aharonov-Bohm Potential]{ Dirac Particle in an Aharonov-Bohm Potential: The Structure of
the First Order S-matrix}
\author{M.S.Shikakhwa\\}
\address{Department of Physics, University of Jordan\\
11942--Amman, Jordan}%
\email{moody@ju.edu.jo}%
 \maketitle

\section{Introduction}
The observation \citep{feinberg,corinaldesi} that the first order
Born amplitude of a non-relativistic spinless particle in an
Aharonov-Bohm (AB) potential \citep{ahar59} does not coincide with
the exact amplitude when the latter is expanded to the same order
triggered an unusual interest in the perturbative aspects of this
problem.  Various remedies for this essentially mathematical
problem were suggested. Examples of some works that addressed this
problem for non-relativistic spinless and spin one-half particles
are
\cite{hagen1,boz1,ouvry,renormalization,ruijinaars,Rumiyyeh,lozano,hagen2,fainberg}.
 For a Dirac particle, the success of the Born amplitude in
providing results consistent with the exact amplitude expanded to
the same order was demonstrated for the first order amplitude in
\cite{vera},and for the second order in \cite{boz2}. A partial
wave calculation of the first order amplitude was also recently
reported \citep{shikakhwa} where consistency with the exact
amplitude was also established. \\
The present work focuses on the interaction Hamiltonian in the
first order $S-$matrix element of a Dirac particle. The algebraic
properties of this interaction Hamiltonian reported
\emph{partially} for the first time in a cursory manner in
\cite{shikakhwa} are exploited in depth and new interesting
properties are reported. In particular, we note the major focus in
the above work is on carrying out partial wave analysis of the
first order Born amplitude of a Dirac particle  in an AB potential
using the basis of the conserved angular momentum operator. In the
present work, on the other hand, the emphasis is not on the
calculations, no amplitudes or so are reported. It is an analysis
of the structure of the first order S-matrix in view of the
algebraic properties of the interaction Hamiltonian. These
analyses which are carried out without using any explicit wave
functions enjoy an intrinsic importance by themselves and provide
a  new insight (as will be shown) into the structure of the first
order transition in the AB potential. Verification of this
structure using explicit wave functions is also given.
\section{Formalism}
A Dirac particle in an electromagnetic field is governed by the
Hamiltonian ($\hbar=c=1$):
\begin{equation}
\label{1}H=H_0+H_{\mathrm{int}}
\end{equation}
where
\begin{equation}
\label{2}H_0=\boldsymbol{\alpha}\cdot\mathbf{p}+\beta m
\end{equation}
and
\begin{equation}
\label{3}H_{\mathrm{int}}=eA_0-e\boldsymbol{\alpha}\cdot\mathbf{A}
\end{equation}
Here, $e$ is the charge of the particle,
${\alpha}_i=\beta\gamma_i$ and $\beta=\gamma_4$. The $\gamma$'s
are the Dirac matrices: $\{\gamma_\mu,\gamma_\nu\}=2g_{\mu\nu}$.
While most of the treatment in this paper is independent of the
explicit representation of these matrices, the Dirac-Pauli
representation is used whenever called for:
\begin{equation}
\label{4} \gamma ^i  = \left( {\begin{array}{*{20}c}
   0 & {\sigma ^i }  \\
   { - \sigma ^i } & 0  \\

 \end{array} } \right), \,\,\,\,\, \gamma ^4  = \left( {\begin{array}{*{20}c}
   I & 0  \\
   0 & { - I}  \\

 \end{array} } \right),
\end{equation}
where $\sigma_i$'s ($i=1,..,3$) are the Pauli matrices, and $I$ is
the $2\times 2$ identity matrix. The first order $S$-matrix
element for the particle is :
\begin{equation}
\label{5}S_{fi}^{(1)}=- i\int {d^4 x\,\bar \psi _f \left( x
\right)\left( {e\gamma _\mu  A^\mu  } \right)\psi _i \left( x
\right)}.
\end{equation}
For the AB-potential \citep{ahar59}, we have:
\begin{equation}
\label{6}A_0=0,
\end{equation}
and
\begin{equation}
\label{7}\mathbf{A}=\frac{\Phi}{2\pi\rho}\hat{\epsilon}_\varphi,
\end{equation}
where $\rho=\sqrt{x^2+y^2}$, $\hat{\epsilon}_\varphi$ is the unit
vector in the $\varphi$-direction, and $\Phi$ is the flux through
the AB tube. The first order $S$-matrix in this case can then be
written in the form:
\begin{equation}
\label{8}S_{fi}^{(1)}=i\alpha \int {d\rho\, d\varphi\,
\psi^{\dagger} _f \left( {\mathbf{x}} \right)\left( {O^ +   + O^ -
} \right)\psi _i \left( {\mathbf{x}} \right)}.
\end{equation}
Here, $\alpha=-e\Phi/2\pi$ (in perturbative calculations
$0<\alpha<1$). The $z$ degree of freedom in the above matrix
element was suppressed, since the $z$-component of the momentum of
the particle-as is well-known-is conserved in the transition. This
simplifies the treatment and does not introduce any loss of
generality. An overall energy conserving $\delta$-function
resulting from the time integration was also suppressed.The
$O^\pm$ operators introduced above are defined as:
\begin{equation}
\label{9} O^\pm \equiv\big( \frac{\alpha_2\pm i\alpha_1}{2}\big)
e^{\pm i\varphi}
\end{equation}
The above operators (note that
$H_{\mathrm{int}}=\frac{O^{+}+O^{-}}{\rho}$)have interesting
properties. One can easily check the following properties
($\Sigma_i=\frac{i}{2}[\gamma_i,\gamma_j] ;\; i,j=1..3$) :
\begin{eqnarray}
  [O^{+},O^{-}] &=& -2(\frac{\Sigma_{3}}{2})\label{10} \\
    \{O^{+},O^{-}\}\equiv O^{+}O^{-}+O^{-}O^{+}&=& I\label{11} \\
  (O^{\pm})^{\dagger} &=& O^{^{\mp}}\label{12} \\
  (O^{\pm})^{2} &=& 0\label{13}
\end{eqnarray}
Moreover, the algebra of $ O^{\pm}$ with the third components of
the spin, $\frac{\Sigma_{3}}{2}$, and the orbital angular
momentum, $ L_{3}$, is also interesting;
\begin{eqnarray}
  [\frac{\Sigma_{3}}{2},O^{\pm}] &=& \mp O^{\pm} \label{14}\\
  \left[L_{3},O^{\pm}\right] &=& \pm O^{\pm} = -[\frac{\Sigma_{3}}{2},O^{\pm}]\label{15}\\
   \{\frac{\Sigma_{3}}{2},O^{\pm}\}&=& 0.\label{16}
\end{eqnarray}
Eqs.(\ref{14}) and  (\ref{15}) suggest that the $O^{\pm}$
operators play the role of raising and lowering operators in the
space of the eigenstates of  $\frac{\Sigma_{3}}{2}$ and $L_{3}$.
Now,  products of the $O^{\pm}$ operators give rise to the
following two Hermitian operators:
\begin{equation}\label{17}
P^{\pm}\equiv O^{\mp}O^{\pm}
\end{equation}
which, in view of Eqs.(\ref{10}) and (\ref{11}) have the explicit
forms
\begin{equation}
\label{18}
 P^{\pm}=\frac{1}{2}\pm \frac{\Sigma_{3}}{2}
\end{equation}
These two operators have a series of interesting properties that
follow from the set of equations, Eqs.( \ref{10}-\ref{13})
\begin{eqnarray}
  (P^{\pm})^{2} &=& P^{\pm} \label{19}\\
   P^{+}P^{-}=P^{-}P^{+}&=&0 \label{20} \\
  P^{+}+P^{-}= \{O^{+},O^{-}\}&=& I.\label{21}
\end{eqnarray}
Eq.(\ref{18}), on the other hand leads to the following
remarkable property of $P^{\pm}$:
\begin{equation}\label{22}
 \frac{\Sigma_{3}}{2}P^{\pm}=\pm\frac{1}{2}P^{\pm}
\end{equation}
The above properties suggest that $P^{\pm}$ is a projection
operator . Indeed, for an arbitrary Dirac spinor $\Psi$ , we have:
\begin{equation}\label{22b}
( P^{+}+P^{-})\Psi=P^{+}\Psi+P^{-}\Psi=\Psi
\end{equation}
with $P^{+}\Psi\quad(P^{-}\Psi)$ being, in view of Eq.(\ref{22}),
an eigenstate of $\frac{\Sigma_{3}}{2}$ with eigenvalue
$+\frac{1}{2}\quad(-\frac{1}{2})$ .Thus, $P^{\pm}$ project out
$\frac{\Sigma_{3}}{2}$ eigenstates with $\pm\frac{1}{2}$
eigenvalues out of an arbitrary $\Psi$, thus allowing the
splitting of this $\Psi$ into a linear combination of these
eigenstates. The following properties of the products of
$P^{\pm}$ and $O^{\pm}$ follow by invoking Eq.(\ref{13}):
\begin{eqnarray}
  P^{+}O^{+} &=& P^{-}O^{-}= O^{+}P^{-} = O^{-}P^{+}=0 \label{24}\\
  O^{+} &=& O^{+}P^{+} = P^{-}O^{+}P^{+}\label{25} \\
  O^{-} &=& O^{-}P^{-} = P^{+}O^{-}P^{-}.\label{26}
\end{eqnarray}
The following brackets can also be easily verified:
\begin{eqnarray}
  [O^{\pm},P^{+}]&=& \mp O^{\pm} \label{27}\\
  \left[O^{\pm},P^{-}\right]&=& \pm O^{\pm}=- [O^{\pm},P^{+}]\label{28}\\
  \left[ \frac{\Sigma_{3}}{2} ,P^{\pm} \right] &=& [L_{3},P^{\pm}]=0.\label{29}
\end{eqnarray}
Before leaving this section, we record the following major result,
which is a consequence of Eqs.(\ref{11}), (\ref{22}),(\ref{25})
and (\ref{26}) :
\begin{equation}\label{30}
\frac{\Sigma_{3}}{2}O^{\pm}=\mp\frac{1}{2}O^{\pm}.
\end{equation}
This relation says that $O^{\pm}$ acting on any arbitrary $\Psi$
project it onto an eigenstate of $ \frac{\Sigma_{3}}{2}$ with
eigenvalues $\mp\frac{1}{2}$.
\section{The First Order $S-$Martrix}
In this section, we will consider the consequences of the
formalism developed above on the first order scattering matrix
element. We start by writing Eq.(\ref{8}) in the form
\begin{equation}\label{31}
S^{(1)}_{fi}=i\alpha M
\end{equation}
where
\begin{eqnarray}
  M&=&\int d\rho d\varphi \psi^{\dagger}_{f}(x)(O^{+}+O^{-})\psi
_{i}(x) \nonumber \\
   &=&  _{f}<\xi'|(O^{+}+O^{-})|\xi>_{i}\label{32}
\end{eqnarray}
In the second line of the above equation we have switched to the
Dirac notation which will be adapted now on . The labels $\xi$ and
$\xi'$ denote the set of all quantum numbers of the initial and
final states, respectively. This includes, for example, the
quantum numbers of the free Hamiltonian $H_{0}$; a spin operator
that commutes with $H_{0}$...etc. This set does not include the
quantum numbers of $\frac{\Sigma_{3}}{2}$, as this does not
commute with the free Hamiltonian and is not a constant of the
transition. The amplitude $M$; Eq.(\ref{32}), with the aid of
Eqs.(\ref{25}) and (\ref{26}) can be written as
\begin{equation}\label{34}
M =
_{f}<\xi'|P^{-}O^{+}P^{+}|\xi>_{i}+_{f}<\xi'|P^{+}O^{-}P^{-}|\xi>_{i}
\end{equation}
which, as a consequence of Eq.(\ref{22}) reduces to
\begin{equation}\label{35}
M =
_{f}<\eta',\frac{-1}{2}|O^{+}|\eta,\frac{+1}{2}>_{i}+_{f}<\eta',\frac{+1}{2}|O^{-}|\eta,\frac{-1}{2}>_{i}
\end{equation}
The states $|\eta,\frac{\pm 1}{2}>_{i}$ and $|\eta',\frac{\pm
1}{2}>_{f}$ ($\pm\frac{1}{2}$ in the above states are the quantum
numbers of $\frac{\Sigma_{3}}{2}$) are,respectively, the
$\frac{\Sigma_{3}}{2}-$components of the initial and final free
particle states. These are projected out of these states through
the action of the operators $P^{\pm}$ in accordance with
Eq.(\ref{22}) . $ \eta$ and $\eta'$ are the set of all the other
quantum numbers of the states, which are generally different from
$ \xi$ and $\xi'$ . The picture drawn by the above equation is
that in the first order matrix element, the two operators
$O^{\pm}$ constituting the effective interaction Hamiltonian split
both the incident and outgoing waves into their
$\frac{\Sigma_{3}}{2}-$components ( the two eigenstates of
$\frac{\Sigma_{3}}{2}$) . The transition, is then the sum of two
transitions induced by $O^{+}$ and $O^{-}$, with $O^{+}$ ($O^{-}$)
linking only the $\frac{\Sigma_{3}}{2}-$component of the incident
wave with eigenvalue $+\frac{1}{2}$ ($-\frac{1}{2}$) to the
$\frac{\Sigma_{3}}{2}-$component of the outgoing wave with eigenvalue $-\frac{1}{2}$ ($+\frac{1}{2}$).\\
 Eq.(\ref{30}) , on the other hand,
tells us that the states $ O^{\pm}|\eta,\pm\frac{1}{2}>_{i}$ are
eigenstates of $\frac{\Sigma_{3}}{2}$ with eigenvalues
$\mp\frac{1}{2}$. This means that the  $O^{\pm}$ operators flip
the spin of the $\frac{\Sigma_{3}}{2}$ states. Thus, we can define
new states
\begin{equation}\label{36}
\widehat{\left|\eta,\mp\frac{1}{2}\right>}_{i}\equiv
O^{\pm}\left|\eta,\pm\frac{1}{2}\right>_{i}
\end{equation}
( note the change $\pm\frac{1}{2}\rightarrow  \mp\frac{1}{2}$ in
the  $\frac{\Sigma_{3}}{2}$ quantum number). We then write the
transition matrix $M$ as
\begin{equation}\label{37}
M =
_{f}\left<\eta',\frac{-1}{2}\right|\widehat{\left.\eta,\frac{-1}{2}\right>}_{i}+_{f}\left<\eta',\frac{+1}{2}\right.\widehat{\left|\eta,\frac{+1}{2}\right>_{i}}.
\end{equation}
 The picture is even more interesting when one works with partial
waves \citep{shikakhwa}. Here, one expands the incident and
outgoing plane waves of the $S-$matrix in terms of the $J-$waves
that are eigenstates of the set of commuting operators: The total
angular momentum operator $J_{3}=L_{3}+\frac{\Sigma_{3}}{2}$, the
Hamiltonian $H_0$ and the spin operator $S_3\equiv\beta\Sigma_3$;
\begin{equation}\label{45}
\begin{aligned}
  H_0 \left|E,j=l+\frac{1}{2},s\right>&  = E\left|E,j=l+\frac{1}{2},s\right>  \\
  J_3 \left|E,j=l+\frac{1}{2},s\right>&  = j_3 \left|E,j=l+\frac{1}{2},s\right> = \left( {\ell  + \tfrac{1}
{2}} \right)\left|E,j=l+\frac{1}{2},s\right> , \,\,\, \ell  = 0, \pm 1,.... \\
  S_3 \left|E,j=l+\frac{1}{2},s\right>&  = s\left|E,j=l+\frac{1}{2},s\right> , \,\,\, s =  \pm 1,
\end{aligned}
\end{equation}
The $S-$matrix is then given as a sum over the partial amplitudes
 $M_{l}$ ( see section \ref{sec 4}
below) :
\begin{equation}\label{38}
M_{l}=\left
<E,j=l+\frac{1}{2},s\right|O^{+}+O^{-}\left|E,j=l+\frac{1}{2},s\right>
\end{equation}
The conserved $J_{3}$ quantum number $j$ is set to
$l+\frac{1}{2}$, $ l=0\pm 1,\pm 2,...$. The quantum numbers $s$
also survives the transition as well as the corresponding operator
is also conserved. The above $J-$waves which are energy
eigenstates as well, are not eigenstates of $\frac{\Sigma_{3}}{2}$
nor of $L_{3}$. Eqs.(\ref{25}) and (\ref{26}), along with
Eq.(\ref{22}) have the following consequences:

\begin{eqnarray}
  O^{+}\left|E,j=l+\frac{1}{2},s\right> &=& P^{-}O^{+}\left|\lambda,j=l+\frac{1}{2},s;l,+\frac{1}{2}\right>\label{39}\\
  O^{-}\left|E,j=l+\frac{1}{2},s\right>&=& P^{+}O^{-}\left|\lambda,j=l+\frac{1}{2},s;l+1,-\frac{1}{2}\right>\label{40}
\end{eqnarray}
The states on the R.H.S are the components of the $J-$waves that
are eigenstates of the operators $L_{3}$ and
$\frac{\Sigma_{3}}{2}$ with eigenvalues $l$, $l+1$ and $\pm
\frac{1}{2}$, respectively. We denote these with the $L-$waves.
$\lambda$ is a collective index for any other possible quantum
number; energy is not one of these , however. The quantum numbers
of $\frac{\Sigma_{3}}{2}$ and
 $L_{3}$ in the above $L-$waves have been fixed by the use of Eq.(\ref{22b}), the fact that $j$ is conserved and
 fixed to $l+\frac{1}{2}$, and that
 $[J_{3},L_{3}]=[S_{3},L_{3}]=[J_{3},\frac{\Sigma_{3}}{2}]=[S_{3},\frac{\Sigma_{3}}{2}]=0$. Plugging
 Eqs.(\ref{39}) and (\ref{40}) into (\ref{38}) , and allowing $P^{\pm}$ to act to the left, we
 get
 \begin{eqnarray}\label{41}
    M_{l}&=&\left
<\lambda',j=l+\frac{1}{2},s;
l+1,-\frac{1}{2}\right|O^{+}\left|\lambda,j=l+\frac{1}{2},s;l,+\frac{1}{2}\right>\nonumber\\
&      &+\left <\lambda',j=l+\frac{1}{2},s;
l,+\frac{1}{2}\right|O^{-}\left|\lambda,j=l+\frac{1}{2},s;l+1,-\frac{1}{2}\right>.
\end{eqnarray}
The incident and outgoing $J-$waves are split by the interaction
into their $L-$wave components, with $O^+$ ($O^-$) linking
incident $L-$waves with the set of eigenvalues $l,+\frac{1}{2}$
($l+1,-\frac{1}{2}$) to outgoing $L-$waves with the set of
eigenvalues $l+1,-\frac{1}{2}$ ($l,+\frac{1}{2}$). Note that the
action of $O^{\pm}$ on the initial $L-$waves leads to raising or
lowering of the quantum numbers of the operators $L_3$ and
$\frac{\Sigma_{3}}{2}$ as dictated by Eqs.(\ref{14}) and
(\ref{15}) . Note also that this takes place in a manner that
conserves the total angular momentum quantum number
$j=l+\frac{1}{2}$. So, we can again define new states  :
\begin{eqnarray}
  \widehat{\left|\xi,j=l+\frac{1}{2},s;l+1,-\frac{1}{2}\right>} &\equiv& O^{+}\left|\xi,j=l+\frac{1}{2},s;l,+\frac{1}{2}\right> \label{43}\\
   \widehat{\left|\xi,j=l+\frac{1}{2},s;l,+\frac{1}{2}\right>}&\equiv&
   O^{-}\left|\xi,j=l+\frac{1}{2},s;l+1,-\frac{1}{2}\right>.\label{44}
\end{eqnarray}

\section{The Matrix element Using Explicit Wave Functions}\label{sec 4}
 In this section, explicit partial wave
functions are going to be used to verify the results of the
previous section. The incident and outgoing wave functions in the
matrix element, Eq.(\ref{8}), are expanded in terms of the eigen
functions of the conserved total angular momentum operator
$J_{3}$. These functions ( $J-$waves) that are  simultaneous eigen
functions of the set of operators $H_{0}$, $J_{3}$ and
$S_{3}=\beta\Sigma_{3}$ in accordance with Eq.(\ref{45}) have the
explicit form \citep {shikakhwa}:
\begin{equation}\label{46}
     \psi _{\ell s} \left( {\mathbf{x}} \right) =
\frac{{e^{i\ell \varphi } }} {{\sqrt {2\pi } \sqrt {2E} \sqrt {2s}
}}\left( {\begin{array}{*{20}c}
   {\sqrt {E + sm} \sqrt {s + 1} J_\ell  \left( \xi  \right)}  \\
   {i\epsilon _3 e^{i\varphi } \sqrt {E - sm} \sqrt {s - 1} J_{\ell  + 1} \left( \xi  \right)}  \\
   {\epsilon _3 \sqrt {E + sm} \sqrt {s - 1} J_\ell  \left( \xi  \right)}  \\
   {ie^{i\varphi } \sqrt {E - sm} \sqrt {s + 1} J_{\ell  + 1} \left( \xi  \right)}  \\

 \end{array} } \right)
\end{equation}
Here $J_\ell\left(\xi\right)$ is the Bessel function of order
$\ell$ ($\xi=p_\bot \rho$), and
$\epsilon_3=\operatorname{sgn}(s)$. $p_\bot$ is the magnitude of
the planar momentum. The above eigen functions are normalized as
\begin{equation}
\label{47} \int {d\rho \,d\varphi \,\psi _{\ell 's'}^\dag  \left(
{\mathbf{x}} \right)\psi _{\ell s} \left( {\mathbf{x}} \right) =
\frac{1} {{p_ \bot  }}\delta \left( {p_ \bot   - p'_ \bot  }
\right)\delta _{\ell \ell '} \delta _{ss'} } .
\end{equation}
If one takes $\psi _i \left( {\mathbf{x}} \right)$ and $\psi _f
\left( {\mathbf{x}} \right)$ to be eigenstates of $S_{3}$ as well,
then one can verify the following expansion of these in terms of
the  $\psi_{\ell s}(\mathbf{x})$ functions \citep{shikakhwa}:
\begin{equation}
\label{48}
\begin{aligned}
  \psi _i \left( {\mathbf{x}} \right)& = \sqrt {E_i } \sum\limits_\ell  {(i)^\ell  \psi _{\ell s} \left( {\mathbf{x}} \right)}  \hfill \\
  \psi _f \left( {\mathbf{x}} \right)& = \sqrt {E_f } \sum\limits_\ell  {(i)^\ell  e^{ - i\ell \theta } \psi _{\ell s} \left( {\mathbf{x}} \right)} . \hfill \\
\end{aligned}
\end{equation}
where $\theta$ is the scattering angle. So, the first order
$S-$matrix element, Eq.(\ref{8}), takes the form
\begin{equation}
\label{49}S_{fi}^{(1)}  = i\alpha E\sum\limits_\ell  {(i)^\ell
\sum\limits_{\ell '} {( - i)^{\ell '} e^{i\ell '\theta } }
}M_{\ell}
\end{equation}
where $M_{\ell}$ is now given by
\begin{equation}
\label{50}
  M_{\ell} = \int {d\rho \,d\varphi \, \psi _{\ell 's'}^{\dagger} \left( {\mathbf{x}} \right)\left( {O^ +   + O^ -  } \right)\psi _{\ell s} \left( {\mathbf{x}} \right)}
\end{equation}
with $O^{\pm}$ as defined earlier. It is easy to check now that
\begin{eqnarray}
\label{51}
  \left( P^{+}+P^{-}\right)\psi _{\ell s} \left( {\mathbf{x}} \right)=\psi _{\ell s} \left( {\mathbf{x}} \right) &=& P^{+}\psi _{\ell s} \left( {\mathbf{x}} \right) +P^{-}\psi _{\ell s} \left( {\mathbf{x}} \right)\nonumber\\
  &=&\phi_{\ell ,+\frac{1}{2}} \left( {\mathbf{x}} \right)+  \phi_{\ell+1 ,-\frac{1}{2}} \left( {\mathbf{x}} \right)
\end{eqnarray}
where the functions
\begin{eqnarray}
  \phi _{\ell,+\frac{1}{2}} \left( {\mathbf{x}} \right)& =& \frac{1}
{{\sqrt {2\pi } \sqrt {2E} \sqrt {2s} }} {\left(
{\begin{array}{*{20}c}
   {\sqrt {E + sm} \sqrt {s + 1} J_\ell  \left( \xi  \right)e^{i\ell \varphi } }  \\
   0  \\
   {\epsilon _3 \sqrt {E + sm} \sqrt {s - 1} J_\ell  \left( \xi  \right)e^{i\ell \varphi } }  \\
   0  \\
 \end{array} } \right)}\label{52}\\
   \phi _{\ell+1,-\frac{1}{2}} \left( {\mathbf{x}} \right)&=& \frac{1}{{\sqrt {2\pi } \sqrt {2E} \sqrt {2s} }}{  \left( {\begin{array}{*{20}c}
   0  \\
   {i\epsilon _3 \sqrt {E - sm} \sqrt {s - 1} J_{\ell  + 1} \left( \xi  \right)e^{i\left( {\ell  + 1} \right)\varphi } }  \\
   0  \\
   {i\sqrt {E - sm} \sqrt {s + 1} J_{\ell  + 1} \left( \xi  \right)e^{i\left( {\ell  + 1} \right)\varphi } }  \\
 \end{array} } \right)}\label{53}
\end{eqnarray}
 are easily checked to be eigenstates of $\frac{\Sigma{3}}{2}$ and $L_{3}$
-as they should be- with eigenvalues $+\frac{1}{2}$
($-\frac{1}{2}$) and $l$ ($l+1$) , respectively. Thus, with the
use of Eqs.(\ref{24}-\ref{26}) we have now $M_{\ell}$ as:
\begin{equation}\label{54}
M_{\ell}=\int d\rho d\varphi\left(\phi
_{\ell+1,-\frac{1}{2}}^{\dagger}\left( {\mathbf{x}}
\right)O^{+}\phi _{\ell,+\frac{1}{2}} \left( {\mathbf{x}}
\right)+\phi _{\ell,+\frac{1}{2}}^{\dagger} \left( {\mathbf{x}}
\right)O^{-}\phi _{\ell+1,-\frac{1}{2}} \left( {\mathbf{x}}
\right)\right)
\end{equation}
which is just Eq.(\ref{41}). Note that the primes on the quantum
numbers have been dropped as the matrix element is diagonal in
these quantum numbers since the corresponding operators $J_{3}$
and $S_{3}=\beta\Sigma_{3}$ are conserved. Now, following
Eqs.(\ref{43}) and (\ref{44}), we define
\begin{eqnarray}
 \widehat{ \phi} _{\ell+1,-\frac{1}{2}} \left( {\mathbf{x}} \right)=O^{+}\phi _{\ell,+\frac{1}{2}} \left( {\mathbf{x}} \right) &=& \frac{1}
{{\sqrt {2\pi } \sqrt {2E} \sqrt {2s} }} {\left(
{\begin{array}{*{20}c}
   0  \\
   {i\epsilon _3 \sqrt {E + sm} \sqrt {s - 1} J_{\ell} \left( \xi  \right)e^{i\left( {\ell  + 1} \right)\varphi } }  \\
   0  \\
   {i\sqrt {E + sm} \sqrt {s + 1} J_{\ell} \left( \xi  \right)e^{i\left( {\ell  + 1} \right)\varphi } }
 \end{array} } \right)} \label{56}\nonumber\\
   \widehat{\phi} _{\ell,+\frac{1}{2}} \left( {\mathbf{x}} \right)=O^{-} \phi _{\ell+1,-\frac{1}{2}} \left( {\mathbf{x}} \right)
   &=& \frac{1}{{\sqrt {2\pi } \sqrt {2E} \sqrt {2s} }}{  \left( {\begin{array}{*{20}c}
  {\sqrt {E - sm} \sqrt {s + 1} J_{\ell+1}  \left( \xi  \right)e^{i\ell \varphi } }  \\
   0  \\
   {\epsilon _3 \sqrt {E - sm} \sqrt {s - 1} J_{\ell+1}  \left( \xi  \right)e^{i\ell \varphi } }  \\
   0 \nonumber \\
 \end{array} } \right)} \label{55}\\
\end{eqnarray}
It is a trivial task to check that $\widehat{ \phi}
_{\ell+1,-\frac{1}{2}} \left( {\mathbf{x}} \right)$ and
$\widehat{\phi} _{\ell,+\frac{1}{2}} \left( {\mathbf{x}} \right)$
are eigenstates of $\frac{\Sigma{3}}{2}$ and $L_{3}$ with the
indicated eigenvalues. At this point, with the explicit wave
functions at hand, it would be interesting to investigate the
relation between the $L-$waves
 $\widehat{\phi} _{\ell,+\frac{1}{2}} \left( {\mathbf{x}}
\right)$ ($\widehat{\phi} _{\ell+1,-\frac{1}{2}} \left(
{\mathbf{x}} \right)$) and the incident $L-$waves $\phi
_{\ell,+\frac{1}{2}} \left( {\mathbf{x}} \right)$ ($\phi
_{\ell+1,-\frac{1}{2}} \left( {\mathbf{x}} \right)$). Comparing
Eqs.(\ref{52}) and  (\ref{56}), for example, we see that
$\widehat{\phi} _{\ell,+\frac{1}{2}} \left( {\mathbf{x}} \right)$
is -apart from a constant- the same as $\phi _{\ell,+\frac{1}{2}}
\left( {\mathbf{x}} \right)$ except for the order of the Bessel
function which is shifted by $+1$ in the former. It is, on the
other hand, shifted by $-1$ in $\widehat{\phi}
_{\ell+1,-\frac{1}{2}} \left( {\mathbf{x}} \right)$ in comparison
to $\phi _{\ell+1,-\frac{1}{2}} \left( {\mathbf{x}} \right)$. This
shift in the order of the Bessel function gives rise to the phase
shift of each partial amplitude. It was shown in \cite{shikakhwa}
that the integral in the partial amplitudes $M_\ell$ reduces to an
integral over Bessel functions of the form $\int d\rho J
_{\ell+1}J _{\ell}$, and this integral gives the partial phase
shifts.  Thus, The operators $O^{\pm}$, while flipping the spin
and orbital angular momentum quantum numbers of each ingoing
$L-$wave to match that of the outgoing one, keeps the order of the
Bessel functions in the two waves shifted by $\pm1$ so as to
generate the the phase shifts in the scattering process. To put
this on more formal grounds, we define the two operators
$d^{\pm}$:
\begin{equation}\label{60}
d^{\pm}\equiv\left(\partial_{\rho}\pm\frac{i}{\rho}\partial_{\varphi}\right)=\left(\partial_{\rho}\mp\frac{L_{3}}{\rho}\right)
\end{equation}
that satisfy
\begin{equation}\label{61}
\left[d^{\pm},L_{3}\right]=\left[d^{\pm},\frac{\Sigma_{3}}{2}\right]=0.
\end{equation}
We note that these operators - upon employing the  well-known
recurrence relations of Bessel functions- have the property
\begin{eqnarray}
  d^{+}J_\ell\left( \xi  \right)e^{i\ell \varphi } &=& -p_\bot J_{\ell+1}  \left( \xi  \right)e^{i\ell \varphi } \\
  d^{-}J_{\ell+1}\left( \xi  \right)e^{i(\ell+1)\varphi }  &=&  p_\bot J_{\ell}  \left( \xi  \right)e^{i(\ell+1) \varphi
  }
\end{eqnarray}
 Therefore, we can now write:
\begin{eqnarray}
 \widehat{ \phi} _{\ell+1,-\frac{1}{2}} \left( {\mathbf{x}} \right)
 =O^{+}\phi _{\ell,+\frac{1}{2}} \left( {\mathbf{x}}
 \right)&=&\frac{1}{p_\bot}\sqrt{\frac{E+sm}{E-sm}}d^{-} \phi _{\ell+1,-\frac{1}{2}} \left( {\mathbf{x}}
 \right)\label{58}\\
  \widehat{ \phi} _{\ell,+\frac{1}{2}} \left( {\mathbf{x}} \right)
 =O^{-}\phi _{\ell+1,-\frac{1}{2}} \left( {\mathbf{x}}
 \right)&=&-\frac{1}{p_\bot}\sqrt{\frac{E-sm}{E+sm}}d^{+} \phi _{\ell,+\frac{1}{2}} \left( {\mathbf{x}}
 \right)\label{59}
\end{eqnarray}
Substituting the above two equations into Eq.(\ref{54}), we get :
\begin{equation}\label{57}
M_\ell=\frac{1}{p_\bot}\int d\rho
d\varphi\left[\sqrt{\frac{E+sm}{E-sm}}\left(\phi
_{\ell+1,-\frac{1}{2}}^{\dagger}\left( {\mathbf{x}} \right)d^{-}
\phi _{\ell+1,-\frac{1}{2}} \left( {\mathbf{x}}
 \right)\right)-\sqrt{\frac{E-sm}{E+sm}}\left(\phi _{\ell,+\frac{1}{2}}^{\dagger} \left(
{\mathbf{x}} \right)d^{+} \phi _{\ell,+\frac{1}{2}} \left(
{\mathbf{x}}
 \right)\right)\right]
\end{equation}
The above equation says that we can write each partial amplitude
as a sum of two transitions, with each one being a transition
among two $L-$waves that have the same $L_3$ and
$\frac{\Sigma_3}{2}$ quantum numbers. These $L-$waves are
connected by the $d^\pm$ operators which merely affect the radial
component of each partial wave, thus inducing the partial phase
shifts. It is interesting indeed that in each of the two terms of
the amplitude, the quantum numbers of the orbital angular momentum
$L_3$  and the spin $\frac{\Sigma_3}{2}$ are conserved in the
transition. It is as if we have replaced the $L_3$  and
$\frac{\Sigma_3}{2}$ non-conserving interactions $O^\pm$ with the
$d^\pm$ interactions that conserve these quantum numbers at the
level of each partial $L-$wave.
\section{Concluding Remarks}
A new insight into the structure of the first order $S$-matrix of
a Dirac particle in an AB potential was provided through the
derivation  of interesting  algebraic properties of the
interaction Hamiltonian appearing in the matrix element. It was
shown that the interaction Hamiltonian is the sum of the two
operators $O^{\pm}$ and algebraic properties of these and related
operators were reported.  The fact that these operators have the
properties (Eqs.(\ref{25},\ref{26})) $O^{\pm}=
P^{\mp}O^{\pm}P^{\pm}$
 along with the fact that $\frac{\Sigma_3}{2}P^\pm=\pm\frac{1}{2}P^\pm$ (Eq.(\ref{22})), were shown to suggest
  the following picture of the transition matrix element:
 The interaction splits both the
incident and outgoing plane waves into their (unconserved)
$\frac{\Sigma_3}{2}$ components so that the matrix element is the
sum of two transitions from
 two initial eigenstates of the  $\frac{\Sigma_3}{2}$ operator to  final eigenstates
 of the same operators .  Each of the operators $O^{\pm}$,
by flipping the spin as a result of the property (Eq.(\ref{30}))
$\frac{\Sigma_{3}}{2}O^{\pm}=\mp\frac{1}{2}O^{\pm}$,links only the
states with opposite spin quantum numbers.The transition, then,
becomes the sum of two transitions induced by the two operators ,
each taking place in one of the spin channels.  At the level of
partial waves, the transition in each partial amplitude is the sum
of two transitions taking place among the two unconserved $L-$wave
components of each incident and outgoing $J-$wave.The fact that
the total angular momentum of each partial wave is conserved leads
to a flip not only of the $\frac{\Sigma_{3}}{2}$ quantum number ,
but also of the $L_3$ quantum number in the transition. Working
with explicit partial wave functions, we have shown that it is
possible to write each partial amplitude as the sum of two
transitions, induced by a couple of two new operators, $d^\pm$,
taking place among $L-$partial wave functions such that the
$\frac{\Sigma_{3}}{2}$ and $L_3$ quantum numbers are conserved in
each of these
transitions. \\
Finally, note that in deriving Eq.(\ref{35}), no assumptions were
made regarding the spin operator that one chooses to diagonalize
with the free Hamiltonian. In the case of partial wave analysis
(Eq.(\ref{41})), however, the spin operator was taken to be
$S_3=\beta\Sigma_3$. The result, however, depends only on the fact
that this spin operator commutes with the operators $O^\pm$, i.e
is a constant of the transition. Any other spin operator having
this property will lead to the same result.

\textbf{Acknowledgments} \\
The author is indebted to professor N.K.Pak and
Dr. B.Tekin for a critical reading of an earlier version of the
manuscript, and for very helpful discussions. He also thanks the
physics department of Middle East Technical University at Ankara
where this work was initiated for hospitality. Partial financial
support by The Scientific and Technical Research Council of Turkey
(T\"{U}B\'{I}TAK) is also acknowledged.

\end{document}